\documentclass[10pt,a4paper]{article}
\usepackage{multicol}
\usepackage{graphicx}
\usepackage{amsmath}
\usepackage[IL2]{fontenc}
\usepackage{color}                  % used only for the \todo{} command
\usepackage{xspace}
\usepackage[ruled,linesnumbered,noline,noend]{algorithm2e}
\usepackage{tikz}
\usetikzlibrary{shapes,arrows}
\usepackage{url}                    % format URLs
\usepackage{listings}               % format code
\usepackage[colorlinks=true,allcolors=blue,breaklinks,draft=false]{hyperref}   % hyperlinks, including DOIs and URLs in bibliography

\newcommand{\todo}[1]{{\color{red}~#1~}}
\newcommand{\honza}[1]{{\color{purple}~#1~}}
\newcommand{\bluetext}[1]{{\color{blue}~#1~}}

\renewcommand{\todo}[1]{}
\renewcommand{\honza}[1]{}
\renewcommand{\bluetext}[1]{}

\newcommand{\Looperman}{\textsc{Looperman}\xspace}
\newcommand{\Loopus}{\textsc{Loopus}\xspace}
\newcommand{\KoAT}{\textsc{KoAT}\xspace}
\newcommand{\PUBS}{\textsc{PUBS}\xspace}
\newcommand{\Rank}{\textsc{Rank}\xspace}

\newcommand{\var}[1]{\texttt{#1}}
\newcommand{\sym}[1]{\ensuremath{\underline{#1}}}
\newcommand{\false}{\ensuremath{\mathit{false}}}
\newcommand{\true}{\ensuremath{\mathit{true}}}
\newcommand{\ite}{\ensuremath{\mathbf{ite}}}
\newcommand{\maxim}{\ensuremath{\mathbf{max}}}
\newcommand{\minim}{\ensuremath{\mathbf{min}}}

\newcommand{\expr}{\mathit{expr}}

\DontPrintSemicolon
\newcommand{\aarg}[2]{#1~{\it //~#2}}
\newcommand{\aargm}[2]{\\ \qquad \aarg{#1}{#2}}
\newcommand{\aset}{\ensuremath{\longleftarrow}}

\begin{document}

%{{{ Title + Authors

\title{Tighter Loop Bound Analysis\\
{\large (Technical report)}
  % Counter-based Symbolic Loop Bound Analysis
  % Symbolic Counter-based Loop Bound Analysis
  % Symbolic Counter-based WCET Analysis
}

\author{
 Pavel \v{C}adek\\
 \textit{Faculty of Informatics}\\
 \textit{Vienna University of Technology}\\
 \textit{Austria}
 \and
 Jan Strej\v{c}ek\\
 \textit{Faculty of Informatics}\\
 \textit{Masaryk University, Brno}\\
 \textit{Czech Republic}
 \and
 Marek Trt\'{\i}k\\
 \textit{LaBRI}\\
 \textit{University of Bordeaux}\\
 \textit{France}
}

%}}}

\date{}

\maketitle

%\twocolumn[
%\begin{@twocolumnfalse}
%\maketitle
%\end{@twocolumnfalse}
%]
\pagenumbering{arabic}

%{{{ Abstract

\begin{abstract}
  We present a new algorithm for computing upper bounds on the number
  of executions of each program instruction during any single program
  run. The upper bounds are expressed as functions of program input values.
  The algorithm is primarily designed to produce bounds
  that are relatively tight, i.e.~not unnecessarily blown up. % The
  % algorithms presents a novel utilization of a loop summarisation
  % technique, which is that was originally suggested for faster
  % symbolic execution.
  %
  The upper bounds for instructions allow us to infer loop bounds,
  i.e.~upper bounds on the number of loop iterations. Experimental
  results show that the algorithm implemented in a prototype tool
  \Looperman often produces tighter bounds than current tools for loop
  bound analysis.
\end{abstract}


%}}}

%{{{ Introduction

\section{Introduction}\label{sec:intro}

The goal of \emph{loop bound analysis} is to derive for each loop in a given
program an upper bound on the number of its iterations during any execution
of the program. These bounds can be parametrized by the program input. The
loop bound analysis is an active research area with two prominent
applications: \emph{program complexity} analysis and \emph{worst case
  execution time} (WCET) analysis.

The aim of program complexity analysis is to derive an asymptotic complexity of
a given program. The complexity is commonly considered by programmers in their
everyday work and it is also used in specifications of programming languages,
e.g.~every implementation of the standard template library of C++ has to have
the prescribed complexities. Loop bound analysis clearly plays a crucial role in
program complexity analysis. In this context, emphasis is put on large coverage
of the loop bound analysis (i.e.~it should find some bounds for as many program
loops as possible), while there are only limited requirements on tightness of
the bounds as asymptotic complexity is studied.

A typical application scenario for WCET analysis is to check whether a
given part of some critical system finishes its execution within an
allocated time budget.  One step of the decision process is to compute
loop bounds. Tightness of the bounds is very important here as an
untight bound can lead to a spuriously negative answer of the analysis
(i.e.~`the allocated time budged can be exceeded'), which may imply
unnecessary additional costs, e.g.~for system redesign or for hardware
components with higher performance.
The WCET analysis can also be used by schedulers to estimate the
run-time of individual tasks.

% If the bounds produced by WCET analysis are tight (and especially when
% the bounds are parametrized by input values), the analysis can be used
% by schedulers to estimate the runtime of individual tasks. 

The problem to infer loop bounds has recently been refined into the
\emph{reachability-bound problem}~\cite{SPEED2010}, where the goal is
to find an upper bound on the number of executions of a given program
instruction during any single run of a given program. One typically
asks for a reachability bound on some resource demanding instruction
like memory allocation. Reachability bound analysis is more
challenging than loop bound analysis as, in order to get a reasonably
precise bound, branching inside loops must be taken into account.

This paper presents a new algorithm that infers reachability bounds.
%for all instructions of a given program. 
More precisely, for each instruction of a given program, the algorithm
tries to find an upper bound on the number of executions of the
instruction in any single run of the program. The bounds are
parametrized by the program input. The reachability bounds can be
directly used to infer loop bounds and asymptotic program complexity.
Our algorithm builds on \emph{symbolic execution}~\cite{Kin76} and
\emph{loop summarisation} %with iteration counters
adopted from~\cite{ST12}.
In comparison with
other techniques for reachability bound or loop bound analysis, our
algorithm brings the following features:
\begin{itemize}
\item It utilizes a loop summarisation technique that computes precise
  values of program variables as functions of loop iteration counts.
\item It distinguishes different branches inside loops and computes
  bounds for each of them separately.
\item If more different bounds arise, it handles all of them while
  other techniques usually choose nondeterministically one of them.
\item It can detect logarithmic bounds.
\item Upper bounds for nested loops are computed more precisely: while
  other techniques typically multiply a bound for the outer loop by a
  maximal bound on iterations of the inner loop during one iteration
  of the outer loop, we sum the bounds for the inner loop over all
  iterations of the outer loop.
  % we sum the bounds for iterations of the inner loop over iterations
  % of the outer loop while other techniques typically multiply the
  % maximal bound for the inner loop by a bound for the outer loop.
  %
  % \item It can discover when a loop has 0 iterations where other
  %   techniques fail. \todo{tomuhle nerozumim.}
\end{itemize}
All these features have a positive effect on tightness of produced bounds.

\begin{figure}[t] %example
\centering
\begin{minipage}[b]{0.4\linewidth}
  \texttt{void f(int x) \{}\\
  \textcolor{white}{aaa}\texttt{int i=5;}\\
  \textcolor{white}{aaa}\texttt{while (i<x)}\\
  \textcolor{white}{aaaaaa}\texttt{i=i+2;}\\%[3ex]
  \texttt{\}}
\end{minipage}
~~~~
\tikzstyle{start} = [regular polygon,regular polygon sides=3, regular polygon rotate=180,thick,draw,inner sep=1.5pt]
\tikzstyle{target} = [regular polygon,regular polygon sides=3, thick,draw,inner sep=1pt]
\tikzstyle{loc} = [circle,thick,draw,minimum size=6mm]
\tikzstyle{pre} = [<-,shorten <=1pt,>=stealth',semithick]
\tikzstyle{post} = [->,shorten <=1pt,>=stealth',semithick]
\begin{tikzpicture}[node distance=1.25cm]
  \node [start,fill=black!10,regular polygon rotate=270] (a) {$a$};
  \node [loc] (b) [right of=a,xshift=.5cm] {$b$}
    edge [pre] node [overlay,label=below:\texttt{i:=5}] {} (a);
  \node [loc] (c) [above of=b] {$c$}
    edge [pre,bend right=60] node [label=above left:\texttt{i<x\!}] {} (b)
    edge [post,bend left=60] node [label=above right:\texttt{\!i:=i+2}] {} (b);
  \node [target,fill=black!10,regular polygon rotate=90] (d)
    [right of=b,xshift=.5cm] {$d$}
    edge [pre] node [overlay,label=below:\texttt{i>=x}] {} (b);
\end{tikzpicture}  
% ~~~~~~~ % verze na vysku
% \begin{tikzpicture}[node distance=1.25cm]
%   \node [start,fill=black!10] (a) {$a$};
%   \node [loc] (b) [below of=a] {$b$}
%     edge [pre] node [overlay,label=left:\texttt{i:=5}] {} (a);
%   \node [loc] (c) [right of=b] {$c$}
%     edge [pre,bend right=60] node [label=above:\texttt{i<x}] {} (b)
%     edge [post,bend left=60] node [label=below:\texttt{~~i:=i+2}] {} (b);
%   \node [target,fill=black!10] (d) [below of=b] {$d$}
%     edge [pre] node [overlay,label=left:\texttt{i>=x}] {} (b);
% \end{tikzpicture}  
%~~~~~~~~~~~
%\begin{tikzpicture}[node distance=1.25cm]
%  \node [start,fill=black!10] (b) {$b$};
%  \node [loc] (c) [below of=b] {$c$}
%    edge [pre] node [label=right:\texttt{i<x}] {} (b);
%  \node [target,fill=black!10] (b') [below of=c] {$b'$}
%    edge [pre] node [label=right:\texttt{i:=i+2}] {} (c);
%\end{tikzpicture}
\caption{A simple C program with its flowgraph.}
\label{fig:ex1} 
\end{figure}
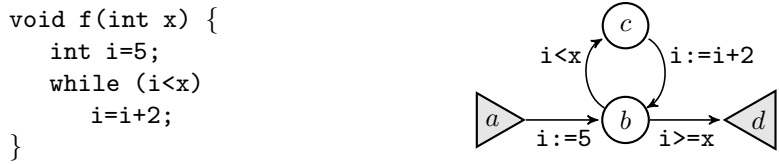

We can explain the basic idea of our algorithm on the
flowgraph %depicted
on the right. The node $a$ is the entry location, $d$ is the exit
location, and locations $b,c$ form a loop. An initial value of
$\var{x}$ represents program input. We symbolically execute each path
in the loop and assign an iteration counter to it. Then we try to
express the effect of arbitrarily many iterations of the loop using
the iteration counters as parameters. The loop in our example has just
one path $bcb$ that increments $\var{i}$ by 2. Hence, the value of
$\var{i}$ after $\kappa$ iterations is $\sym{i}'+2\kappa$, where
$\sym{i}'$ denotes the value of $\var{i}$ before the loop execution
starts. We combine this loop summary with the program state just
before entering the loop, which is obtained by symbolic execution of
the corresponding part of the program. In our example, we get that the
value of $\var{i}$ after $\kappa$ iterations of the loop is
$5+2\kappa$. To enter another iteration, the condition $\texttt{i<x}$
must hold. If we replace the variables $\var{i}$ and $\var{x}$ by
their current values, we get the condition $5+2\kappa<\sym{x}$, where
$\sym{x}$ refers to the initial value of $\var{x}$. This condition is
satisfied only if $\kappa<\frac{\sym{x}-5}{2}$. As $\kappa$ is an
iteration counter, it has to be a non-negative integer. Hence, we get
the bound on the number of loop iterations
$\maxim\{0,\lceil\frac{\sym{x}-5}{2}\rceil\}$, which is assigned to
all edges in the loop. Edges outside the loop are visited at most
once. The situation is more complicated if we have loops with more
loop paths, nested loops, or loops where a run can cycle forever. The
algorithm is described % and illustrated by a more complex example
in Section~\ref{sec:alg}. % preceded by
% Section~\ref{sec:prelim},
% where we define necessary notation and explain basics of symbolic
% execution.

We have implemented our algorithm in an experimental tool \Looperman. Comparison
with several leading loop bound analysis tools shows that our approach often
provides tighter loop bounds. For example, our tool is currently the only one
that detects that the inner cycle of the BubbleSort algorithm makes
$\frac{n\cdot(n-1)}{2}$ iterations in total (i.e.~during all iterations of the
outer loop) when sorting an array of $n$ elements, while other tools provide
only the bound $n^2$ or $\mathcal{O}(n^2)$. Section~\ref{sec:eval} presents the
comparison of \Looperman with four other loop bound analysis tools on 199
benchmarks.

%Related work is discussed in Section~\ref{sec:rw}. Section~\ref{sec:future}
%shows some simple programs for which our algorithm currently does not infer any
%bound. Possible improvements of the algorithm are present at the same time.

%}}}
%{{{ Preliminaries

\section{Preliminaries}\label{sec:prelim}

In this section we introduce or recall some terminology intensively used in
the following sections. For simplicity, this paper focuses on programs
manipulating only integer scalar variables $\var{a},\var{b},\ldots$ and
read-only multidimensional integer array variables $\var{A},\var{B},\ldots$,
and with no function calls.

\subsection{Flowgraph, backbone, loop, induced flowgraph}
\label{sec:flowgraph}

An analysed program is represented as a \emph{flowgraph}
$P=(V,E,l_\mathit{beg},$ $l_\mathit{end},\iota)$, where $(V,E)$ is a finite
oriented graph, $l_\mathit{beg},l_\mathit{end}\in V$ are different
\emph{begin} and \emph{end nodes} respectively. The in-degree of
$l_\mathit{beg}$ and the out-degree of $l_\mathit{end}$ are 0, and
$\iota:E\rightarrow\mathcal{I}$ labels each edge $e$ by an
\emph{instruction} $\iota(e)$. The out-degree of all nodes except the end
node is $1$ or $2$. Nodes with out-degree $2$ are called \emph{branching}
nodes. We use two kinds of instructions: an assignment instruction
$\var{a}\texttt{:=}\expr$ for some scalar variable $\var{a}$ and some
program expression $\expr$ over program variables, and an assumption
$\texttt{assume}(\gamma)$ for some quantifier-free formula $\gamma$ over
program variables. Out-edges of any branching node are labelled with
assumptions $\texttt{assume}(\gamma)$ and $\texttt{assume}(\neg\gamma)$ for
some $\gamma$. We often omit the keyword $\texttt{assume}$ in flowgraphs.

A \emph{path} in a flowgraph is a (finite or infinite) sequence $\pi=v_1 v_2
\cdots$ of nodes such that $(v_i, v_{i+1})\in E$ for all $v_i,v_{i+1}$ in the
sequence. Paths are denoted by Greek letters.

A \emph{backbone} in a flowgraph is an acyclic path leading from the begin node
to the end node.

Let $\pi$ be a backbone with a prefix $\alpha v$. There is a \emph{loop} $C$
with a \emph{loop entry} $v$ along $\pi$, if there exists a path $v\beta v$ such
that no node of $\beta$ appears in $\alpha$. The loop $C$ is then the smallest
set containing all nodes of all such paths $v\beta v$.

Each \emph{run} of the program corresponds to a path in the flowgraph
starting at $l_\mathit{beg}$ and such that it is either infinite, or it is
finite and ends in $l_\mathit{end}$.\footnote{We assume that crashes or
  other undefined behaviour of program expressions are prevented by safety
  guards. For example, an expression \texttt{a/b} is guarded by
  $\texttt{assume}(\var{b}\neq 0)$.}  Every run follows some
backbone: it can escape from the backbone in order to perform
one or more iterations in a loop along the backbone, but once the last
iteration in the loop is finished, the execution continues along the
backbone again. We thus talk about a \emph{run along} a backbone. We can
compute the backbone for a finite run corresponding to a path $\rho$ by the
following procedure: If $\rho$ is acyclic, then the backbone is directly
$\rho$. Otherwise, we find the leftmost repeating node in $\rho$, remove the
part of $\rho$ between the first and the last occurrence of this node
(including the last occurrence), and repeat the procedure. In other words,
the backbone arises from $\rho$ by removing all cycles in it. For an
infinite run, we extend the procedure to remove the suffix starting just
after the leftmost node that repeats infinitely often on the run. The
procedure returns an acyclic path that is a prefix of some backbones. We can
associate the run to any backbone with this prefix.

For a loop $C$ with a loop entry $v$ along a backbone $\pi$, a
\emph{flowgraph induced by the loop}, denoted as $P(C,v)$, is derived from
the subgraph of the original flowgraph induced by $C$, where $v$ is marked
as the begin node, a fresh end node $v'$ is added, % and marked as the end node,
and every transition $(u,v)\in E$ leading to $v$ is redirected to $v'$ (we
identify the edge $(u,v')$ with $(u,v)$ in the context of the original
program). Each single iteration (i.e.~a path from $v$ back to $v$ without
visiting $v$ inside) of the loop corresponds to a run of the
induced flowgraph. 
% is replaced by a transition $(u,\iota,v')$.  A backbone in a flowgraph
% induced by a loop is called \emph{loop path}.

Note that program representation by flowgraphs and the above definition of
loops easily handle many features of programming languages like
\texttt{break}, \texttt{continue}, or even program loops constructed using
\texttt{goto} instructions.

\subsection{Symbolic execution}
\label{sec:symexpr}

The basic idea of \emph{symbolic execution}~\cite{Kin76} is to replace
input data of a program by \emph{symbols} representing arbitrary
data. Executed instructions then manipulate symbolic expressions over
the symbols instead of exact values. A \emph{symbolic expression} is
any term of the theory of integers extended with functions
$\maxim$ and $\minim$, rounding functions $ \lceil\cdot\rceil $ and
$ \lfloor\cdot\rfloor $ for constant expressions over reals, and
\begin{itemize}
\item for each scalar variable $\var{a}$, an uninterpreted constant
  $\sym{a}$, which is a \emph{symbol} representing any initial
  (input) value of the variable $\var{a}$,
\item for each array variable $\var{A}$, an uninterpreted function $\sym{A}$
  of the same arity as $\var{A}$, which is a \emph{symbol} representing any
  initial (input) content of the array $\var{A}$,
\item a countable set $\{\kappa_1,\kappa_2,\ldots\}$ % \cup \{K\}
  of artificial variables (not appearing in analysed programs), called
  \emph{path counters} and ranging over non-negative integers,
\item a special symbol $\star$ called \emph{unknown}, and
\item for each formula $\psi$ build over symbolic expressions and two
  symbolic expressions $\phi_1,\phi_2$, a construct $\ite(\psi,\phi_1,\phi_2)$
  of meaning ``\textbf{i}f-\textbf{t}hen-\textbf{e}lse'', that evaluates to
  $\phi_1$ if $\psi$ holds, and to $\phi_2$ otherwise.
\end{itemize}

Let $\psi,\phi$ be symbolic expressions and $x$ be a symbol or a path
counter. By $\psi[x/\phi]$ we denote the expression $\psi$, where all
occurrences of $x$ are simultaneously replaced by $\phi$. We further extend
this notation such that $\psi[x_i/\phi_i\mid i \in I]$ denotes the
expression $\psi$, where all occurrences of $x_i$ are simultaneously
replaced by $\phi_i$, for each $i \in I$. 

We sometimes add the upper index $\vec{\kappa}$ to expression names to
denote that the expressions can contain path counters $\vec{\kappa} =
(\kappa_1,\ldots,\kappa_n)$. We say that an expression is
\emph{$\vec{\kappa}$-free} if it contains no path counters.

Symbolic execution stores variable values in \emph{symbolic memory} and all
executable program paths are uniquely identified by corresponding \emph{path
  conditions}. Here we provide brief descriptions of these
terms. For more information see~\cite{Kin76}.

A \emph{symbolic memory} is a function $\theta$ assigning to each scalar
variable $\var{a}$ a symbolic expression and to each array variable
$\var{A}$ the symbol $\sym{A}$ (the value of every array variable is always
identical to its initial value as we consider programs with read-only
arrays). A symbolic memory $\theta$ is called \emph{initial}, if
$\theta(\var{a})=\sym{a}$, for each scalar or array variable $\var{a}$.

We overload the notation $\theta(\cdot)$ to program expressions as follows.
Let $\textit{expr}$ be a program expression over program variables
$\var{a}_1$, \ldots, $\var{a}_n$. Then $\theta(\textit{expr})$ represents
a symbolic expression obtained from $\textit{expr}$ such that we
simultaneously replace all occurrences of the variables $\var{a}_1$,
\ldots, $\var{a}_n$ by symbolic expressions $\theta(\var{a}_1)$, \ldots,
$\theta(\var{a}_n)$ respectively.

Symbolic execution of a path in a flowgraph starts with the initial symbolic
memory and the memory is updated on assignments. For example, if the first
executed assignment is \texttt{a:=2*a+b}, the initial symbolic memory
$\theta$ is updated to the symbolic memory $\theta'$ where
$\theta'(\var{a})=\theta(\texttt{2*a+b})=2\sym{a}+\sym{b}$. If we later
update $\theta'$ on \texttt{a:=1-a}, we get the memory $\theta''$ such that
$\theta''(\var{a})=\theta'(\texttt{1-a})=1-(2\sym{a}+\sym{b})=
1-2\sym{a}-\sym{b}$.

If $\psi$ is a symbolic expression over symbols $\{\,\sym{a}_i \mid i \in
I\}$ corresponding to program variables $\{\,\var{a}_i \mid i \in I\}$
respectively, then $\theta\langle \psi \rangle$ denotes the symbolic
expression $ \psi[\sym{a}_i/\theta(\var{a}_i) \mid i \in I] $. For example,
if $\theta(\var{a})=\kappa_1$ and $\theta(\var{b})=\sym{a}-\kappa_2$, then
$\theta\langle2\sym{a}+\sym{b}\rangle=2\theta(\var{a})+\theta(\var{b})=
2\kappa_1+\sym{a}-\kappa_2$. We apply the notation
$\theta\langle\varphi\rangle$ and $\varphi[x/\psi]$ with the analogous
meanings also to formulae $\varphi$ built over symbolic
expressions. 

Note that $\theta_1\langle\theta_2(\var{a})\rangle$ returns the value of
$\var{a}$ after a code with effect $\theta_1$ followed by a code with effect
$\theta_2$. For example, if $\theta_1(\var{a})=2\sym{a}+1$ represents the
effect of assignment \texttt{a:=2*a+1} and $\theta_2(\var{a})=\sym{a}-2$ the
effect of assignment \texttt{a:=a-2}, then
$\theta_1\langle\theta_2(\var{a})\rangle=\theta_1\langle\sym{a}-2\rangle
=(2\sym{a}+1)-2$ represents the effect of the two assignments applied in the
sequence.

Given a path in a flowgraph leading from its begin node $l_\mathit{beg}$,
the \emph{path condition} is a formula over symbols, which determines the
set of all possible program inputs for which the program execution will
follow the path. A path condition is constructed during symbolic execution
of the path. Initially, the path condition is set to \textit{true} and it
can only be updated when an $\texttt{assume}(\gamma)$ is executed. For
example, if a symbolic execution reaches $\texttt{assume(a > 5)}$ with a
path condition $\varphi$ and a symbolic memory $\theta(\var{a}) =
2\sym{a}-1$, then it updates the path condition to
$\varphi\,\wedge\,2\sym{a}-1 > 5$.

\subsection{Upper bound}
\label{sec:upperbound}

An \emph{upper bound} for an edge $e$ in a flowgraph $P$ is a
$\vec{\kappa}$-free symbolic expression $\rho$ % containing no $\star$
such that whenever $P$ is executed on any input, the instruction on edge $e$
is executed at most $\rho'$ times, where $\rho'$ is the expression that we
get by replacing each variable symbol by the input value of the
corresponding variable.

%}}}
%{{{ The Algorithm

\section{The Algorithm}\label{sec:alg}

Recall that every program run follows some backbone and the run can diverge
from the backbone only to loops along the backbone. The algorithm first
detects all backbones. For each backbone $\pi_i$ and each edge $e$, it
computes a set of upper bounds $\beta_i(e)$ on the number of visits of the
edge by any run following the considered backbone. As all these bounds are
valid, the set $\beta_i(e)$ can be interpreted as a single bound
$\minim\,\beta_i(e)$ on visits of edge $e$ by any run along $\pi_i$. At the
end, the overall upper bound for an edge $e$ can be computed as the maximum of
these bounds over all backbones, i.e.~as
$\maxim\{\minim\,\beta_i(e)\mid \pi_i~\text{ is a backbone}\}$.
 
The algorithm consists of the following four procedures:
\begin{description}
\item[\texttt{executeProgram}] is the starting procedure of the whole
  algorithm. It gets a flowgraph and computes all its backbones. Then it
  symbolically executes each backbone and computes for each edge a set of
  upper bounds on the number of visits of the edge by a run following the
  backbone. Whenever the symbolic execution visits a loop entry node, the
  procedure \texttt{processLoop} is called to get upper bounds on visits
  during loop execution.
\item[\texttt{processLoop}] gets a loop represented by the program induced
  by the loop. Note that each run of the induced program corresponds to one
  iteration of the loop and it follows some backbone of the induced program
  (the backbones are called \emph{loop paths} in this context). The
  procedure then symbolically executes each loop path by recursive call of
  \texttt{executeProgram} (the nesting of recursive calls thus directly
  corresponds to the nesting of loops in the program). The recursive call of
  \texttt{executeProgram} produces, for each loop path, a symbolic memory
  and a path condition capturing the effect of a single iteration along the
  loop path. The procedure \texttt{processLoop} then calls
  \texttt{computeSummary}, which takes the symbolic memories after single
  loop iterations, assigns to each loop path a unique path counter
  $\kappa_i$, and computes a \emph{parametrized symbolic memory}
  $\theta^{\vec{\kappa}}$ describing the effect of an arbitrary number of
  loop iterations. This symbolic memory is parametrized by path counters
  $\vec{\kappa}=(\kappa_1,\ldots,\kappa_k)$ representing the numbers of
  iterations along the corresponding loop paths. From the parametrized
  symbolic memory and from the path conditions corresponding to single loop
  iterations (received from the recursive call of \texttt{executeProgram}),
  we derive a \emph{parametrized necessary condition} for each loop path,
  which is a formula over symbols and path counters $\vec{\kappa}$ that has
  to be satisfied when another loop iteration along the corresponding loop
  path can be performed after $\vec{\kappa}$ loop iterations. Finally,
  \texttt{processLoop} infers upper bounds from these parametrized necessary
  conditions with the help of the procedure \texttt{computeBounds}.
\item[\texttt{computeSummary}] is a subroutine of \texttt{processLoop} that
  gets symbolic memories corresponding to one loop iteration along each loop
  path and it produces the parametrized symbolic memory
  $\theta^{\vec{\kappa}}$ after an arbitrary number of loop iterations (as
  mentioned above).
\item[\texttt{computeBounds}] is another subroutine of
  \texttt{processLoop}. It gets a set $I$ of loop paths and the corresponding
  parame\-tri\-zed necessary conditions and tries to derive some upper bounds on
  the number of loop iterations along loop paths from $I$.
\end{description}
% We store bounds for an edge $e$ as a set $S$ of simple expressions rather than
% one complex bound. For example, if we have two bounds for one edge, say
% $\maxim\{0,\sym{n}{-}2\}$ and $\maxim\{0,\sym{x}{+}\sym{y}\}$, we store them
% as the set $\{\maxim\{0,\sym{n}{-}2\},\maxim\{0,\sym{x}{+}\sym{y}\}\}$ instead
% of a single bound
% $\minim\{\maxim\{0,\sym{n}{-}2\},\maxim\{0,\sym{x}{+}\sym{y}\}\}$. The simple
% strucuture of bounds plays an important role when the \texttt{processLoop}
% procedure infers bounds for edges laying on a nested loop (we point out the
% reason in Section~\ref{ssec:processLoop}).

We describe the four procedures in the following four subsections. The
procedure \texttt{processLoop} is described as the last one as it calls the
other three procedures.

  %{{{ executeProgram

\subsection{Algorithm \Large \texttt{executeProgram}}

\begin{algorithm}[!t]
\small
\caption{
\texttt{executeProgram}
%\texttt{executeProgram(}$P$\texttt{)}
}
\label{alg:executeProgram}
\KwIn{
    \aargm{$P$~}
    {a flowgraph}
}
\KwOut{
    {\\\qquad\!\!\! $\{(\pi_1,\theta_1,\varphi_1),\ldots,(\pi_k,\theta_k,\varphi_k)\}$
    {\it // backbones $\pi_i$ (with symbolic memory $\theta_i$ 
    \\\hspace{4.9cm}// and path condition $\varphi_i$ after execution \\\hspace{4.9cm}// along~$\pi_i$)}}
% abstract symbolic states \\\hspace{1.1cm}// $\theta_i,\varphi_i$ after execution along $\pi_i$}}
    \aargm{$\beta$~}
%    {upper bounds for edges in the flowgraph}
    {for each edge $e$ of $P$, $\beta(e)$ is a set of upper bounds for $e$}
}
\BlankLine 
$\var{states}\aset\emptyset$\;
Compute the set of backbones $\{\pi_1,\ldots,\pi_k\}$ of $P$.\;
\ForEach{$i=1,\ldots,k$} { \label{alg:executeProgram:ForEachBackbone}
  Initialize $\theta_i$ to return $\sym{a}$ for each (scalar or array) variable $\var{a}$.\;
  $\varphi_i\aset\true$\;  
  Initialize $\beta_i$ to return $\{0\}$ for each edge.\;
  Let $\pi_i=v_1 \ldots v_n$.\;
  \ForEach{$j=1,\ldots,n-1$} { \label{alg:executeProgram:NodeOfBackbone}
    \If{$v_j$ is a loop entry} { \label{alg:executeProgram:LoopEntry}
      Let $C$ be the loop with the loop entry $v_j$ along $\pi_i$.\;
      % Compute induced flowgraph $P(C,v_j)$.\;
      $(\beta_\mathit{loop},\theta_i)\aset 
        \texttt{processLoop(}P(C,v_j),\theta_i,\varphi_i\texttt{)}$\;
      \label{alg:executeProgram:callProcessLoop}
      \ForEach{edge $e$ of $P(C,v_j)$} {
        \label{alg:executeProgram:addBounds}
        $\beta_i(e)\aset
        \{\rho_1+\rho_2\mid\rho_1\in\beta_i(e),\,\rho_2 \in \beta_\mathit{loop}(e)\}$\;
      }
    }
    \If {$\iota((v_j, v_{j+1}))$ has the form {\tt assume(}$\psi${\tt )} 
        and $\theta_i(\psi)$ contains no $\star$} {
        \label{alg:executeProgram:assume}
      $\varphi_i\aset\varphi_i\wedge\theta_i(\psi)$\;
      % \If {$\varphi_i$ is not satisfiable} {
      %     \label{alg:executeProgram:unsat}
      %     Continue at line~\ref{alg:executeProgram:insert}.
      % }
    }
    \If {$\iota((v_j, v_{j+1}))$ has the form $\var{\tt a}:=\expr$} {
      $\theta_i(\var{a})\aset\theta_i(\expr)$\;
    }
    $\beta_i((v_j, v_{j+1}))\aset\{\rho+1\mid\rho\in\beta_i((v_j,v_{j+1}))\}$\;
    \label{alg:executeProgram:increment}
  }
  Insert $(\pi_i,\theta_i,\varphi_i)$ into $\var{states}$.\;
  \label{alg:executeProgram:insert}
}
\ForEach{edge $e$ of $P$} { \label{alg:executeProgram:Merge}
  $\beta(e)\aset\{\maxim\{\rho_1,\dots,\rho_k\}\mid\rho_1\in\beta_1(e),
  \ldots,\,\rho_k\in\beta_k(e)\}$\;
}
\Return{$(\var{\tt states},\beta)$}
\end{algorithm}

The procedure \texttt{executeProgram} of Algorithm~\ref{alg:executeProgram}
takes a flowgraph as input, determines its backbones, and symbolically
executes each backbone separately. For a backbone $\pi_i$, symbolic
execution computes symbolic memory $\theta_i$, path condition $\varphi_i$,
and \emph{bound function} $\beta_i$ assigning to each edge $e$ a set of
symbolic expressions that are valid upper bounds on the number of visits of
edge $e$ during any single run along the backbone. Each such a set
$\beta_i(e)$ of bounds could be replaced by a single bound
$\minim~\beta_i(e)$, but we prefer to keep it as a set of simpler
expressions to increase the success rate of expression matching in procedure
\texttt{processLoop} (we point out the reason in
Section~\ref{ssec:processLoop}).

The symbolic execution proceeds in the standard way until we enter a loop
entry (line~\ref{alg:executeProgram:LoopEntry}). Then we call procedure
\texttt{processLoop} on the loop, current symbolic memory and path
condition. The procedure returns function $\beta_\mathit{loop}$ of upper
bounds on visits of loop edges during execution of the loop, and a symbolic
memory after execution of the loop. We add these bounds and the former
bounds in the \textbf{foreach} loop at
line~\ref{alg:executeProgram:addBounds} and continue the execution along the
backbone.  If the \texttt{processLoop} procedure cannot determine the value
of some variable after the loop,
%(possibly because the value depends on the number of iterations)
it simply uses the symbol $\star$ (unknown). 

Another difference from the standard symbolic execution is at
line~\ref{alg:executeProgram:assume} where we suppress insertion of
predicates containing $\star$ to the path condition. As a consequence, a
path condition of our approach is no longer a necessary and sufficient
condition on input values to lead the program execution along the
corresponding path (which is the case in standard symbolic execution), but
it is only a necessary condition on input values of a run to follow the
backbone.

% Once the path condition becomes unsatisfiable, we stop the execution of the
% backbone (line~\ref{alg:executeProgram:unsat}) as there is no run further
% following this backbone (there can be an infinite run following some prefix
% of this backbone). 

After processing an edge of the backbone, we increase the corresponding
bounds by one (line~\ref{alg:executeProgram:increment}).

At the end of the procedure, the resulting bounds for each edge are computed
as the maximum of previously computed bounds for the edge over all backbones
(see the \textbf{foreach} loop at
line~\ref{alg:executeProgram:Merge}). Besides these bounds, the procedure
also returns each backbone with the symbolic memory and path condition after
its execution.
 
  %}}}
  %{{{ computeSummary

\subsection{Algorithm \Large \texttt{computeSummary}}

\begin{algorithm}[!t]
\small
\caption{
\texttt{computeSummary}
%\texttt{computeSummary(}$\{(\pi_1,\theta_1), \ldots, (\pi_l,\theta_l)\}$\texttt{)}
}
\label{alg:compSum}
\KwIn{ 
  {\\\qquad $\{(\pi_1,\theta_1), \ldots, (\pi_l,\theta_l)\}$
  {\it // each $ \theta_i $ is a memory after a single 
    execution of $\pi_i$}}
}
\KwOut{
  \aargm{$\theta^{\vec{\kappa}}$~}{the symbolic memory after $\vec{\kappa}$
    iterations of backbones $\pi_1,\ldots,\pi_l$}
}
\BlankLine

Introduce fresh path counters $\vec{\kappa}=(\kappa_1,\ldots,\kappa_l)$ for backbones $\pi_1,\ldots,\pi_l$ resp.\; \label{alg:compSum:elim:end}
Initialize $\theta^{\vec{\kappa}}$ to return $\star$ for each scalar
variable, and $\sym{A}$ for each array variable $\var{A}$.\;
\Repeat{$\var{\tt change} = \false$}{ \label{alg:compSum:repeat}
  $\var{change} \aset \false$\; 
  \label{alg:compSum:initByStars}
  
  \ForEach{variable $\var{a}$ such that 
      $\theta^{\vec{\kappa}}(\var{a})=\star$}{ 
    \label{alg:compSum:foreach}
    \label{l:monotFirstLoop}
    Compute an improved value $b$ for the variable \var{a} from symbolic 
    memories $\theta_1,\ldots,\theta_l$ and $\theta^{\vec{\kappa}}$.\; 
    \label{alg:compSum:improveVar}
    \If{$b\ne\star$}{
            \label{alg:compSum:canImprove}
      $\theta^{\vec{\kappa}}(\var{a})\aset b$\;
            \label{alg:compSum:improve}
      $\var{change} \aset \true$\;
    }
  }
}
\Return{$\theta^{\vec{\kappa}}$}\;
\end{algorithm}

The procedure \texttt{computeSummary} of Algorithm~\ref{alg:compSum} is a
subpart of the procedure of the same name introduced in~\cite{ST12}.

The procedure gets loop paths $\pi_1,\ldots,\pi_l$ together with
symbolic memories $\theta_1,\ldots,\theta_l$, where each $\theta_i$
represents the effect of a single iteration along $\pi_i$. Then it
assigns fresh path counters $\vec{\kappa}=(\kappa_1,\ldots,\kappa_l)$
to the loop paths and computes the parametrized symbolic memory
$\theta^{\vec{\kappa}}$ after $\vec{\kappa}$ iterations of the loop,
i.e.~after $\sum_{1\le i\le l}\kappa_i$
%$\kappa_1+\kappa_2+\ldots+\kappa_l$
iterations where
exactly $\kappa_i$ iterations follow $\pi_i$ for each $i$ and there is
no assumption on the order of iterations along different loop paths.

Note that the value of some variables can depend on the order of iterations
along different loop paths. If we do not find the precise value of some
variable after $\vec{\kappa}$ iterations, then $\theta^{\vec{\kappa}}$
assigns $\star$ (unknown) to this variable.

To be on safe side, we start with $\theta^{\vec{\kappa}}$ assigning $\star$
to all scalar variables. Then we gradually improve the precision of
$\theta^{\vec{\kappa}}$ as long as there is some progress. The crucial step
is the computation of an improved value $b$ for a scalar variable $\var{a}$
at line~\ref{alg:compSum:improveVar}. The value $b$ is defined as $\star$
except the following four cases.
\begin{enumerate}
\item For each loop path $\pi_i$, we have $\theta_i(\var{a})=\sym{a}$. In
  other words, the value of $\var{a}$ is not changed in any iteration of the
  loop. This case is trivial. We set $b=\sym{a}$.
\item For each loop path $\pi_i$, either $\theta_i(\var{a})=\sym{a}$ or
  $\theta_i(\var{a})=\sym{a}+d_i$ (resp. $\theta_i(\var{a})=\sym{a}\cdot d_i$)
  for some symbolic expression $d_i$ such
  that $\theta^{\vec{\kappa}}\langle d_i\rangle$ contains neither $\star$
  nor any path counters. Let us assume that the latter possibility holds for
  loop paths $\pi_1,\ldots,\pi_{m}$. The condition on
  $\theta^{\vec{\kappa}}\langle d_i\rangle$ guarantees that the value of
  $d_i$ is constant during all iterations over the loop. In this case, we
  set $b=\sym{a}+\sum_{i=1}^{m}\theta^{\vec{\kappa}}\langle d_i\rangle\cdot\kappa_i$
  (resp. $b=\sym{a}\cdot\prod_{i=1}^{m}\theta^{\vec{\kappa}}\langle d_i\rangle^{\kappa_i}$).
\item There exists a symbolic expression $d$ such that
  $\theta^{\vec{\kappa}}\langle d\rangle$ contains neither $\star$ nor any
  path counters. For each loop path $\pi_i$, either
  $\theta_i(\var{a})=\sym{a}$ or $\theta_i(\var{a})=d$. Let us assume that
  the latter possibility holds for loop paths $\pi_1,\ldots,\pi_{m}$.  In
  other words, the value of $\var{a}$ is set to $d$ in each iteration along
  loop path $\pi_i$ for $1\le i\le m$, while it is unchanged in any other
  iteration. Hence, we set $b=\ite(\sum_{1\le i\le
    m}\kappa_i>0,\theta^{\vec{\kappa}}\langle d\rangle,\sym{a})$.
\item For one loop path, say $\pi_i$, $\theta_i(\var{a})=d$ for some
  symbolic expression $d$ such that $\theta^{\vec{\kappa}}\langle d\rangle$
  contains neither $\star$ nor any path counters except $\kappa_i$. Further,
  for each loop path $\pi_j$ such that $i\neq j$,
  $\theta_j(\var{a})=\sym{a}$. That is, only iterations along $\pi_i$ modify
  $\var{a}$ and they set it to a value independent on other path counters
  than $\kappa_i$. Note that if we assign $d$ to $\var{a}$ in the
  $\kappa_i$-th iteration along $\pi_i$, then the actual assigned value of
  $d$ is the value after $\kappa_i-1$ iterations along the paths. Therefore
  we set $b=\ite(\kappa_i>0,(\theta^{\vec{\kappa}}\langle
  d\rangle)[\kappa_i/\kappa_i-1],\sym{a})$.
\end{enumerate}
Note that one can add another cases covering other situations where the
value of $\var{a}$ can be expressed precisely, e.g.~the case capturing
geometric progressions.  Also note that
%  symbolic value $b$ computed by the
% rules represent precise value for the analysed variable
% $\var{a}$. Therefore, 
the value of any variable in the resulting symbolic memory
$\theta^{\vec{\kappa}}$ is expressed either precisely, or it is 
unknown (i.e.~$\star$).
% We may consider abstract values in between the corner cases, but we leave
% this for future research.

  %}}}
  %{{{ computeBounds

\subsection{Algorithm \Large \texttt{computeBounds}}
 
\begin{algorithm}[!t]
\small
\caption{
\texttt{computeBounds}
%\texttt{computeBounds(}$I,\varphi$\texttt{)}
}
\label{alg:computeBounds}
\KwIn{
    \aargm{$I$~}
    {indices of backbones}
    \aargm{$\varphi$~}
    {a necessary condition to perform an iteration along a backbone\\ 
    \hspace{1.1cm}// with an index in $I$ after $\vec{\kappa}$ iterations}
}
\KwOut{
    \aargm{$B$~}
    {upper bounds on the number of iterations\\ \hspace{1.1cm}//
     along backbones with indices in $I$}
}
\BlankLine
\lIf {$\varphi[\kappa_i / 0 \mid i\in I]$ is not satisfiable} {
  \Return{$\{0\}$}
  \label{alg:computeBounds:specialCase}
}
$B\aset\emptyset$\;
\ForEach{\rm inequality $\sum_{j\in J\supseteq I} a_j\kappa_j < b$ implied by $\varphi$,
  where each $a_j$ is a positive integer and $b$ is $\vec{\kappa}$-free} { 
  \label{alg:computeBounds:inequalities}
  $B\aset B\cup\{\maxim\{0, \lceil b / \minim\{a_i \mid i \in I\}\rceil\}\}$ 
  \label{alg:computeBounds:bound}
}
%Derive a finite set $S$ of inequalities implied by $\varphi$.\;
%\ForEach{inequality of $S$ transformable to the form $\sum_{j\in J\supseteq
%    I} a_j\kappa_j < b$, where each
%  $a_j$ is a positive integer and $b$ is $\vec{\kappa}$-free} { 
%  \label{alg:computeBounds:inequalities}
%  $B\aset B\cup\{\maxim\{0, \lceil b / \minim\{a_i \mid i \in I\}\rceil\}\}$ 
%}
\Return{$B$}
\end{algorithm}

The procedure \texttt{computeBounds} of Algorithm~\ref{alg:computeBounds}
gets a set $I$ of selected loop path indices, and a necessary condition
$\varphi$ to perform an iteration along some loop path with an index in $I$
(we talk about an \emph{iteration along $I$} for short) after $\vec{\kappa}$
previous loop iterations. From this information, the procedure infers upper
bounds on the number of loop iterations along~$I$.

We would like to find a tight upper bound, i.e.~a $\vec{\kappa}$-free
symbolic expression $B$ such that there exist some values of symbols (given
by a valuation function $v$) for which the necessary condition
$\varphi[\,\sym{a}/v(\sym{a})\mid \sym{a}\text{ is a symbol}\,]$ to make another
iteration along $I$ is satisfiable whenever the number of finished
iterations along $I$ is less than $B[\,\sym{a}/v(\sym{a})\mid \sym{a}\text{ is
  a symbol}\,]$ and the same does not hold for the expression $B+1$. An
effective algorithm computing these tight bounds is an interesting research
topic itself.

The presented procedure infers some bounds only for two special cases.
Line~\ref{alg:computeBounds:specialCase} covers the case when even the first
iteration along any loop path in $I$ is not possible: the procedure then
returns the bound $0$.

The other special case is the situation when the necessary condition
implies an inequality of the form $\sum_{j\in J\supseteq I}a_j\kappa_j<b$,
where each $a_j$ is a positive integer and $b$ is $\vec{\kappa}$-free. To
detect these cases, we transform the necessary condition to the conjunctive
normal form and look for clauses that contain just one predicate and try to 
transfer the predicate into this form by basic arithmetical
operations knowing that $\varphi$ holds. Each such inequality implies the following:
\begin{align*}
\sum_{j\in J\supseteq I}a_j\kappa_j<b &\implies \sum_{i\in I}a_i\kappa_i<b
    \implies  \minim\{a_i\mid i\in I\}\cdot\sum_{i\in I}\kappa_i<b.
\end{align*}
Hence, $\sum_{i \in I}\kappa_i<\lceil b/\minim\{a_i\mid i\in I\}\rceil$ has
to be satisfied to perform another iteration along $I$ after $\vec{\kappa}$
previous iterations including $\sum_{i \in I}\kappa_i$ iterations along
$I$. As all path counters are non-negative integers, we derive the bound
$\maxim\{0,\lceil b/\minim\{a_i\mid i\in I\}\rceil\}$ on iterations along
$I$.

  %}}}
  %{{{ processLoop

\subsection{Algorithm \Large \texttt{processLoop}}\label{ssec:processLoop}

\begin{algorithm}[!t]
\small
\caption{
\texttt{processLoop}
%\texttt{processLoop(}$Q,\theta_\mathit{in},\varphi_\mathit{in}$\texttt{)}
}
\label{alg:processLoop}
\KwIn{
      \aargm{$Q$~~~}
      {a flowgraph induced by a loop}
      \aargm{$\theta_{in}$~\,}
      {a symbolic memory when entering the loop}
      \aargm{$\varphi_{in}$~}
      {a path condition when entering the loop}
}
\KwOut{
      \aargm{$\beta_{\mathit{loop}}$}
      {upper bounds for all edges in the loop} % during all its iterations
      \aargm{$\theta_{out}$~}{symbolic memory after the loop}
}
\BlankLine
Initialize $\beta_{\mathit{loop}}$ to return $\emptyset$ for each edge $e$
  of $Q$.\; 
\label{alg:processLoop:part1Begin}
$(\{(\pi_1,\theta_1,\varphi_1),\ldots,(\pi_k,\theta_k,\varphi_k)\},
  \beta_{\mathit{inner}})\aset$ $\texttt{executeProgram(}Q\texttt{)}$\;
\label{alg:processLoop:recCall}
$\theta^{\vec{\kappa}}\aset \texttt{computeSummary(}\{ 
  (\pi_1,\theta_1), \ldots,(\pi_k,\theta_k)\}\texttt{)}$\;
\label{alg:processLoop:callSummary}
$\varphi_i^{\vec{\kappa}}\aset\varphi_\mathit{in}\wedge
  \theta_\mathit{in}\langle\theta^{\vec{\kappa}}\langle\varphi_i\rangle\rangle$ 
  for each $i\in \{1,\ldots,k\}$\;
\label{alg:processLoop:iteratePCs}
$\beta^{\vec{\kappa}}(e)\aset
  \{\theta_\mathit{in}\langle\theta^{\vec{\kappa}}\langle\rho\rangle\rangle 
  \mid\rho\in\beta_{\mathit{inner}}(e)\}$ for each edge $e$ of $Q$\;
\label{alg:processLoop:iterateBounds}
\label{alg:processLoop:part1End}
\ForEach{\rm edge $e$ of $Q$}{
  \label{alg:processLoop:part2Begin}
  $I\aset\{i \mid e \text{ is on } \pi_i \text{ or on a loop along }\pi_i\}$\;
  \label{alg:processLoop:part2:I}
  $B_{\mathit{outer}}\aset
    \texttt{computeBounds(}I,\bigvee_{i \in I}\varphi_i^{\vec{\kappa}}\texttt{)}$\;
  \label{alg:processLoop:part2:B}
  \If{$0\in B_{\mathit{outer}}$}{
    $\beta_{\mathit{loop}}(e)\aset\{0\}$
    \label{alg:processLoop:part2:zero}
  }
  \Else{
    \ForEach{$\rho_{\mathit{inner}}\in \beta^{\vec{\kappa}}(e)$}{
      \If{\rm $\rho_{\mathit{inner}}\equiv c$ where $c$ is $\vec{\kappa}$-free}{
        \label{alg:processLoop:part2:simpleCase}
        $\beta_{\mathit{loop}}(e)\aset\beta_{\mathit{loop}}(e) 
          \cup\{c\cdot\rho_{\mathit{outer}}\mid\rho_{\mathit{outer}}\in B_{\mathit{outer}}\}$
        \label{alg:processLoop:part2:simpleBounds}
      }
      \ElseIf{\rm $\rho_{\mathit{inner}}\equiv\maxim\{c,b+
        \sum_{i=1}^{k}a_i\kappa_i\}$ where $c,b$ and all $a_i$ are 
          $\vec{\kappa}$-free}{
        \label{alg:processLoop:part2:complexCase}
        $J\aset \{j \mid j \notin I \wedge a_j\ne 0\}$\;
        \label{alg:processLoop:part2:complexCase:J}
        $B_J\aset \texttt{computeBounds(}
          J,\bigvee_{j \in J}\varphi_j^{\vec{\kappa}}\texttt{)}$\; 
        \label{alg:processLoop:part2:complexCase:BJ}
        %   \tcp*{$B_J\aset\{0\}$ if $J=\emptyset$}
        \If{$B_J\neq\emptyset$}{
          $b'\aset b + \maxim\{0,a_j \mid j\in J\} \cdot 
            \minim\,B_J$\;
          \label{alg:processLoop:part2:complexCase:bprime}
          $a\aset\maxim \{a_i\mid i\in I\}$\;
          \label{alg:processLoop:part2:complexCase:a}
          \ForEach{$\rho_{\mathit{outer}} \in B_{\mathit{outer}}$} {
            $\beta_{\mathit{loop}}(e)\aset\beta_{\mathit{loop}}(e)
              \,\cup\, \{\sum_{K = 0}^{\rho_{\mathit{outer}}-1} 
              \maxim\{c, b'+a \cdot K\}\}$
            \label{alg:processLoop:part2End}
          }
        }
      }
    }
  }
}
\ForEach{\rm edge $e$ of $Q$}{
  \label{alg:processLoop:part3Begin}
  \If{\rm an edge $e'$ of $Q$ such that $\beta^{\vec{\kappa}}(e')=\emptyset$
    is reachable from $e$ in $Q$}{
    \label{alg:processLoop:infinite:condition}
    $\beta_{\mathit{loop}}(e)\aset
      \{\rho_1+\rho_2\mid\rho_1\in\beta_{\mathit{loop}}(e),\,
      \rho_2\in\beta^{\vec{\kappa}}(e)$, and $\rho_2$ is $\vec{\kappa}$-free$\}$\;
  }
}
$\theta_\mathit{out}(\var{a})\aset
  \theta_\mathit{in}\langle\theta^{\vec{\kappa}}(\var{a})\rangle$
for each variable $\var{a}$\;
\label{alg:processLoop:part4Begin}
Eliminate $\vec{\kappa}$ from $\theta_\mathit{out}$.\;
% \lForEach{\rm variable \var{a} such that $\theta_\mathit{out}(\var{a})$ is
%   not $\vec{\kappa}$-free}{$\theta_\mathit{out}(\var{a})\aset\star$}
\label{alg:processLoop:part4End}
% \texttt{memoryAfterLoop}$(\theta_\mathit{in},\theta^{\vec{\kappa}})$\;
\Return{$(\beta_{\mathit{loop}}, \theta_\mathit{out})$}
\end{algorithm}

The procedure \texttt{processLoop} of Algorithm~\ref{alg:processLoop} gets a
flowgraph $Q$ representing the body of a loop, i.e.~each run of $Q$
corresponds to one iteration of the original loop. We symbolically execute
$Q$ using the recursive call of \texttt{executeProgram} at
line~\ref{alg:processLoop:recCall}. We obtain all loop paths
$\pi_1,\ldots,\pi_k$ of $Q$ and bounds $\beta_\mathit{inner}$ on visits
of each edge in the loop during any single iteration of the loop. For each
$\pi_i$, we also get symbolic memory $\theta_i$ after one iteration along
$\pi_i$ and a necessary condition $\varphi_i$ to perform this iteration.
% \todo{Neni fakt, ze dostaneme backbony, sym. pameti a nutne podminky,
% zrejmy uz z toho, ze zavolame Alg.1?}
The procedure \texttt{computeSummary} produces the parametrized symbolic
memory $\theta^{\vec{\kappa}}$ after
% any number of loop iterations. More precisely, the symbolic memory is
% parametrized by path counters $\vec{\kappa}=(\kappa_1,\ldots,\kappa_k)$
% and it represents values of program variables after
% $\kappa_1+\ldots+\kappa_k$ iterations (or after
$\vec{\kappa}$ iterations.
% for short) of the loop such that, for each $i$, exactly $\kappa_i$
% iterations go along the loop path $\pi_i$.
Symbols $\sym{a}$ appearing in $\theta^{\vec{\kappa}}$ refer to variable
values before the loop is entered. If we combine $\theta^{\vec{\kappa}}$
with the symbolic memory before entering the loop $\theta_\mathit{in}$, we
get the symbolic memory after execution of the code preceding the loop and
$\vec{\kappa}$ iterations of the loop. We use this combination to derive
necessary conditions $\varphi_i^{\vec{\kappa}}$ to perform another iteration
% \todo{REVIEW 2: are you sure that you still explore an over-approximation of
%   all viable execution paths? (I guess, you are, but some argument might
%   helpful)}
along $\pi_i$ and upper bounds $\beta^{\vec{\kappa}}$ on visits of loop
edges in the next iteration of the loop. % after $\vec{\kappa}$ iterations.
   
The \textbf{foreach} loop at
line~\ref{alg:processLoop:part2Begin} %--\ref{alg:processLoop:part2End}
computes upper bounds for all edges of the processed loop on visits during
all its complete iterations (incomplete iterations when a run cycles in
some nested loop forever are handled later). We already have the bounds
$\beta^{\vec{\kappa}}$ on visits in a single iteration after $\vec{\kappa}$
preceding iterations.
% Note that the bounds are either $\{1\}$ or $\{0\}$ for edges, that are not
% part of a nested loop.
For each edge $e$, we compute the set $I$ of all loop path indices such that
iterations along these loop paths can visit $e$. 
% i.e.~either $e$ lies on loop paths with indices from $I$, or $e$ lies
% within a (nested) loop along them.
The \texttt{computeBounds} procedure at line~\ref{alg:processLoop:part2:B}
takes $I$ and a necessary condition to perform an iteration along $I$ after
$\vec{\kappa}$ iterations and computes bounds $B_\mathit{outer}$ on the
number of iterations along $I$. If there is 0 among these bounds, $e$ cannot
be visited by any complete iteration and the computation for $e$ is
over. Otherwise we try to compute some overall bounds for each bound
$\rho_\mathit{inner}$ on the visits of $e$ during one iteration (after
$\vec{\kappa}$ iterations) separately. If $\rho_{\mathit{inner}}$ is a
$\vec{\kappa}$-free expression
(line~\ref{alg:processLoop:part2:simpleCase}), then it is constant in each
iteration and we simply multiply it with every bound on the number of
iterations along $I$. The situation is more difficult if
$\rho_{\mathit{inner}}$ contains some path counters. We can handle the
frequent case when it has the form
$\maxim\{c,b+\sum_{i=1}^{k}a_i\kappa_i\}$, where $a_1,\ldots,a_k,b,c$ are
$\vec{\kappa}$-free (see line~\ref{alg:processLoop:part2:complexCase} and
note that this is the reason for keeping the bounds simple). First we
get rid of path counters $\kappa_j$ that have some influence on this bound
(i.e.~$a_j\neq 0$), but $e$ cannot be visited by any iteration along loop
path $\pi_j$. Let $J$ be the set of indices of such path counters
(line~\ref{alg:processLoop:part2:complexCase:J}). We try to compute bounds
$B_J$ on the number of iterations along $J$
(line~\ref{alg:processLoop:part2:complexCase:BJ}), which are also the bounds
on $\sum_{j\in J}\kappa_j$. Note that if $J=\emptyset$, we call
\texttt{computeBounds(}$\emptyset,\false$\texttt{)}, which immediately
returns $\{0\}$. If we get some bounds in $B_J$, we can overapproximate
$\sum_{i=1}^{k}a_i\kappa_i$ as follows:
% \begin{eqnarray*}
%   \sum_{i=1}^{k}a_i\kappa_i&~=~&\sum_{j\in J}a_j\kappa_j+\sum_{i\in I}a_i\kappa_i\\
%   % &\le&\maxim\{0,a_j\mid j\in J\}\cdot\sum_{j\in J}\kappa_j
%   % +\maxim\{a_i\mid i\in I\}\cdot\sum_{i\in I}\kappa_i\\
%   &\le&\maxim\{0,a_j\mid j\in J\}\cdot\minim\,B_J
%   +\maxim\{a_i\mid i\in I\}\cdot\sum_{i\in I}\kappa_i\\
% \end{eqnarray*}
% \todo{Nebylo by dobre tady na ten vzorec zmensit pismo, aby se to veslo na
%   radek?}
\begin{align*}
&\sum_{i=1}^{k}a_i\kappa_i = \sum_{j\in J}a_j\kappa_j+\sum_{i\in I}a_i\kappa_i
    \le\\
&\le~\maxim\{0,a_j\mid j\in J\}\cdot\minim\,B_J +\maxim\{a_i\mid i\in I\}\cdot\sum_{i\in I}\kappa_i
\end{align*}
% \todo{Pozorny ctenar by se mohl ptat, proc davame $\maxim\{0,a_j\mid j\in
% J\}\cdot\minim\,B_J$ a ne $\maxim\{ j\mid j\in J\}\cdot\minim\,B_J$. Duvod
% je ten, ze pokud iterace pres J snizuji bound, je lepsi jich udelat 0.}
Using the definitions of $b'$ and $a$ at
lines~\ref{alg:processLoop:part2:complexCase:bprime}--\ref{alg:processLoop:part2:complexCase:a},
we overapproximate the bound $\rho_\mathit{inner}$ on visits of $e$ during
one iteration along $I$ after $\vec{\kappa}$ loop iterations by
$$\rho_\mathit{inner}=\maxim\{c,b+\sum_{i=1}^{k}a_i\kappa_i\}
~\le~\maxim\{c,b'+a\cdot\sum_{i\in I}\kappa_i\}.$$ As $K$-th iteration along
$I$ is preceded by $K-1$ iterations along $I$, the edge $e$ can be visited
at most $\maxim\{c,b'+a\cdot (K-1)\}$ times during $K$-th iteration. For each
bound $\rho_\mathit{outer}$ on the iterations along $I$, we can now compute
the total bound on visits of $e$ as
$\sum_{K=0}^{\rho_{\mathit{outer}}-1}\maxim\{c, b' + a \cdot K\}$.
 
\begin{figure*}[!t]
\centering
\tikzstyle{start} = [regular polygon,regular polygon sides=3,regular polygon rotate=180,
  thick,fill=black!10,draw,inner sep=1.5pt]
\tikzstyle{target} = [regular polygon,regular polygon sides=3,
  thick,fill=black!10,draw,inner sep=1.5pt]
\tikzstyle{loc} = [circle,thick,draw,minimum size=6mm]
\tikzstyle{pre} = [<-,shorten <=1pt,>=stealth',semithick]
\tikzstyle{post} = [->,shorten <=1pt,>=stealth',semithick]

\begin{tabular}{cc}
\multicolumn{2}{c}{
\begin{tabular}{l}
\texttt{int nonzeros(int n, int* A) \{}\\
\texttt{~~int k=0;}\\
\texttt{~~for (int i=0; i<n \&\& k<3; i++)}\\
\texttt{~~~~if (A[i]!=0)}\\
\texttt{~~~~~~k++;}\\
\texttt{~~return k;}\\
\texttt{\}}
\end{tabular}
}
\\
\multicolumn{2}{c}{(a)}
\\ \\
\begin{tabular}{c}
\hspace{-.4cm}
\begin{tikzpicture}[node distance=1.25cm]
 \node [start,regular polygon rotate=270,inner sep=2pt] (b) {$a$};
 \node [loc] (c) [right of=b,node distance=1.5cm] {$b$}
   edge [pre] node [overlay,label=above:\texttt{k:=0}] {} (b);
 \node [loc] (d) [right of=c,node distance=1.5cm] {$c$}
   edge [pre] node [overlay,label=above:\texttt{i:=0}] {} (c);
 \node [loc] (f) [below of=d,node distance=1.35cm] {$d$}
   edge [pre] node [label=left:\texttt{i<n} $\wedge$ \texttt{k<3}\!] {} (d);
 \node [loc] (j) [below of=f, xshift=1cm] {$f$}   
   edge [pre] node [right, pos=0.7] {~\texttt{A(i)=0}} (f)
   edge [post, out=25,in=330,looseness=1.4] node [pos=0.08,right] {~\texttt{i:=i+1}} (d);
 \node [loc] (g) [below of=f, xshift=-1cm] {$e$}
   edge [post] node [label=below:\texttt{k:=k+1}] {} (j)
   edge [pre] node [left, pos=0.7] {\texttt{A(i)!=0}~~} (f);
 \node [target,regular polygon rotate=90,inner sep=1.8pt] (k) [right of=d, node distance=2.7cm] {$g$}
   edge [pre] node [overlay,label=above:{\texttt{i>=n} $\vee$ \texttt{k>=3}}] {} (d);
\end{tikzpicture}
\end{tabular}
&
\begin{tabular}{c}
\hspace{-.4cm}
\begin{tikzpicture}[node distance=1.25cm]
 \node [start,regular polygon rotate=270,inner sep=2pt,color=white,fill=white] (b) {$a$};
 \node [start,inner sep=2.5pt] (d) [right of=b] {$c$}
   edge [pre,color=white] node [overlay,color=white,label=above:{\texttt{k}}] {} (b);
 \node [loc] (f) [below of=d,node distance=1.35cm] {$d$}
   edge [pre] node [label=left:\texttt{i<n} $\wedge$ \texttt{k<3}\!] {} (d);
 \node [loc] (j) [below of=f,xshift=.9cm] {$f$}   
   edge [pre] node [right, pos=0.7] {~\texttt{A(i)=0}} (f);
 \node [loc] (g) [below of=f,xshift=-.9cm] {$e$}
   edge [post] node [label=below:\texttt{k:=k+1}] {} (j)
   edge [pre] node [left, pos=0.7] {\texttt{A(i)!=0}~~} (f);
 \node [target,regular polygon rotate=180,inner sep=1pt] (k) [right of=d,node distance=1.8cm] {$c'$}
   edge [pre,out=270,in=0,looseness=.85] node [pos=0.75,right] {\texttt{i:=i+1}} (j);
\end{tikzpicture}
\end{tabular}
\\
(b) & (c)
\end{tabular}
\caption{C program (a) with the corresponding flowgraph (b), and its
  induced flowgraph $P(\{c,d,e,f\},c)$ (c). }
\label{fig:RunningExample}
\end{figure*}
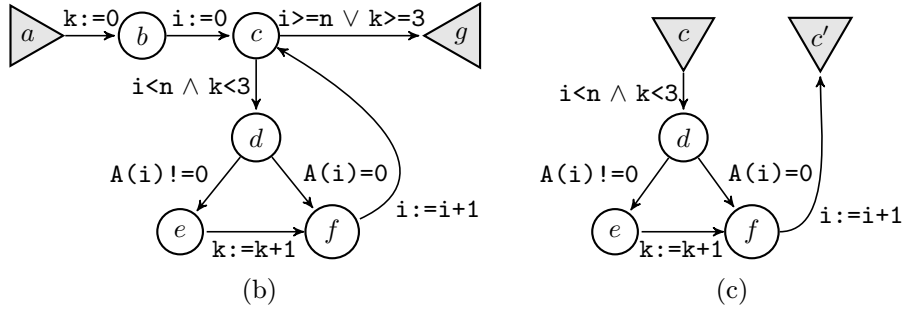

Until now we have considered visits of loop edges during \emph{complete}
iterations. However, it may also happen that an iteration is started, but
never finished because the execution keeps looping forever in some nested
loop. For example, in the program
\texttt{while(x>0)\{x:=x-1;while(true)\{\}\}}, we easily compute bound $0$
on the number of complete iterations of the outer loop and thus we assign
bound $0$ to all loop edges at
line~\ref{alg:processLoop:part2:zero}. However, some edges of the loop are
visited. These incomplete iterations are treated by the \textbf{foreach}
loop at line~\ref{alg:processLoop:part3Begin}. Whenever an edge $e$ can be
visited by an incomplete iteration (which is detected by existence of some
subsequent edge $e'$ without any bound and thus potentially lying on an
infinite nested loop), we add the ($\vec{\kappa}$-free) bounds on visits of
$e$ during one iteration to the total bounds for $e$. If there is no such
$\kappa$-free bound, we leave $e$ unbounded to be on the safe side.

Finally, the lines~\ref{alg:processLoop:part4Begin} and
\ref{alg:processLoop:part4End} combine the symbolic memory before the loop
with the effect of the loop and eliminate loop counters from the resulting
symbolic memory $\theta_\mathit{out}$. Roughly speaking, the elimination
replaces every expression that is not $\vec{\kappa}$-free by $\star$. In
fact, the elimination can be done in a smarter way. For example, after the
loop in the program $\texttt{i:=0;while(i<n)\{i:=i+1\}}$, the elimination
can replace $\kappa$ by $\maxim\{0,\theta_\mathit{out}(\var{n})\}$. 
%\todo{We discuss this extension in the supplement paper.}

  %}}}

%}}}
%{{{ Example

\section{Example}\label{sec:example}

We execute the proposed technique on the example of
Figure~\ref{fig:RunningExample}(a--b). The program accepts an array
$\var{A}$ of size $\var{n}$ and it counts up to three non-zero
elements in the array.

We will observe a run of our algorithm on the flowgraph step by step. We
will follow calls to individual procedures where we present the current
state of the computation in terms of variables of the procedure at the top
of the call stack. In descriptions of symbolic memories we omit variables
that are not modified in the program: the value of $\var{n}$ and $\var{A}$
is always $\sym{n}$ and $\sym{A}$, respectively.

The execution starts by calling \texttt{executeProgram} with the flowgraph at
Figure~\ref{fig:RunningExample}(b). The flowgraph has only one backbone
$\pi_1=\mathit{abcg}$. The node $c$ is the loop entry to the loop
$\{c,d,e,f\}$ along the backbone. Symbolic execution of $\pi_1$ up to $c$
is straightforward and leads to the symbolic memory
$\theta_1(\var{k})=\theta_1(\var{i})=0$, the
path condition $\varphi_1=\true$, and the bound function $\beta_1$ maps each
edge to $\{0\}$ except $\beta_1((a,b))=\beta_1((b,c))=\{1\}$. At the entry node
$ c $ we build an induced flowgraph $ P(\{c,d,e,f\},c) $ depicted in
Figure~\ref{fig:RunningExample}(c). Then we call %\texttt{processLoop}.
\texttt{processLoop(}$P(\{c,d,e,f\},c),$ $\theta_1,\varphi_1$\texttt{)}.

%\texttt{processLoop} needs data from the symbolic execution of the passed
%flowgraph $ P(\{c,d,e,f\},c) $. It thus calls \texttt{executeProgram} on
%this flowgraph.
%
%In this recursive call of \texttt{executeProgram} we are supposed to
%symbolically execute the only backbone $\mathit{cdeic'}$ of the flowgraph $
%P(\{c,d,e,f,g,h,i\},c) $. There is exactly one loop $\{e,f,g,h\}$ at the loop
%entry $e$ along the backbone. The execution up to loop entry $ e $ is
%straightforward. At the entry node we first build the induced flowgraph
%$ P(\{e,f,g,h\},e) $ depicted
%at Figure~\ref{fig:RunningExample}(d) and then we call
%\texttt{processLoop(}$P(\{e,f,g,h\},e), $ $\theta_1,\varphi_1$\texttt{)}, where $
%\theta_1(\var{j}) = 0 $ and $ \varphi_1 = \sym{i} < \sym{n} $.

\texttt{processLoop} calls \texttt{executeProgram} with the
flowgraph at Figure~\ref{fig:RunningExample}(c). Since the flowgraph does not
contain any loop, \texttt{executeProgram} symbolically executes both loop
paths $ \mathit{cdefc'}$ and $ \mathit{cdfc'}$ and returns the result to the
caller. \texttt{processLoop} thus receives backbones $\pi_1 = \mathit{cdefc'}$
and $ \pi_2=\mathit{cdfc'}$, the corresponding symbolic memories and path
conditions
\begin{align*}
\theta_1(\var{i}) &= \sym{i}+1 & \theta_2(\var{i}) &= \sym{i}+1\\
\theta_1(\var{k}) &= \sym{k}+1 & \theta_2(\var{k}) &= \sym{k}\\
\varphi_1 &= \sym{i}<\sym{n} ~\wedge~ \sym{k}<3~\wedge &
\varphi_2 &= \sym{i}<\sym{n} ~\wedge~ \sym{k}<3~\wedge \\
&~~~~~\sym{A}(\sym{i})\neq 0 &
&~~~~~\sym{A}(\sym{i})=0
\end{align*}
and a bound function $\beta_\mathit{inner}$ assigning $\{1\}$ to each edge of
the flowgraph. Now we call \texttt{computeSummary} for the symbolic memories $
\theta_1 $ and $ \theta_2 $ and we get path counters
$\vec{\kappa}=(\kappa_1,\kappa_2)$ for the backbones $ (\pi_1, \pi_2) $
respectively and the following symbolic memory $\theta^{\vec{\kappa}}$
$$
\theta^{\vec{\kappa}}(\var{i})=\sym{i}+\kappa_1+\kappa_2 
\qquad
\theta^{\vec{\kappa}}(\var{k})=\sym{k}+\kappa_1
$$
Next, at line~\ref{alg:processLoop:iteratePCs} we compute necessary conditions
(to perform another iteration along backbones $\pi_1$ and $\pi_2$ respectively)
\[
\begin{array}{rcl}
\varphi^{\vec{\kappa}}_1 & =\, & \kappa_1+\kappa_2<\sym{n}~ 
\wedge~\kappa_1<3~\wedge~\sym{A}(\kappa_1+\kappa_2)\neq0
\\
\varphi^{\vec{\kappa}}_2 & =\, & \kappa_1+\kappa_2<\sym{n}~
\wedge~\kappa_1<3\wedge~\sym{A}(\kappa_1+\kappa_2)=0
\end{array}
\]
The next line produces bound function $\beta^{\vec{\kappa}}$ which is the same
as $\beta_\mathit{inner}$, in this case. Now we have all data we need to start
the computation of resulting bounds for all five edges of the passed flowgraph.

The main part of this computation is performed in the loop at line
\ref{alg:processLoop:part2Begin}. We show the computation for the edge
$ (e,f) $. First we call
\texttt{computeBounds(}$\{1\},
\varphi^{\vec{\kappa}}_1$\texttt{)}. Since the passed formula is
satisfiable for $ \kappa_1 = 0 $, we get to the
line~\ref{alg:computeBounds:inequalities}. The condition there is
satisfied for predicates $ \kappa_1+\kappa_2<\sym{n} $ and
$ \kappa_1 < 3 $. The computation on the subsequent line is
straightforward for both predicates, so we receive the set
$ B_\mathit{outer}=\{\maxim\{0,\sym{n}\}, \maxim\{0,3\} \}
=\{\maxim\{0,\sym{n}\}, 3\} $. Then we get to the
line~\ref{alg:processLoop:part2:simpleBounds} in \texttt{processLoop}
(because $\beta^{\vec{\kappa}}((e,f)) = \{1\}$), where we receive
$\beta_\mathit{loop}((e,f)) = \{\maxim\{0,\sym{n}\}, 3\}$. The
computation proceeds similarly for other edges, but for
$(c,d)$, $(d,f)$, $(f,c)$ it produces only one bound
$\{\maxim\{0,\sym{n}\}\}$. Since, the condition at
line~\ref{alg:processLoop:infinite:condition} is false for all edges,
the resulting $ \beta_\mathit{loop} $ returns
$ \{\maxim\{0,\sym{n}\},$ $ 3\} $ for $(d,e)$ and $(e,f)$, and
$ \{\maxim\{0,\sym{n}\}\} $ for the others. The resulting symbolic
memory $\theta_\mathit{out}$ assigns $\star$ to $\var{i}$ and
$\var{k}$.

The control-flow then returns back to \texttt{executeProgram} where we
update $\beta_1$ according to received $\beta_\mathit{loop}$. Then we
symbolically execute the remaining edge $(c,g)$. The computation in
the loop at line~\ref{alg:executeProgram:Merge} computes maximum over
all bounds for a considered edge.  The algorithm then terminates with
the bound function $\beta$ assigning $\{1\}$ to edges
$(a,b)$, $(b,c)$, $(c,g)$, the set $\{\maxim\{0,\sym{n}\}\}$ to edges
$(c,d)$, $(d,f)$, $(f,c)$, and the set $\{\maxim\{0,\sym{n}\},3\}$ to
$(d,e)$ and $(e,f)$.

We can conclude for the program in Figure~\ref{fig:RunningExample}(a)
that the loop can be executed only if the program is called with some
positive integer $ \sym{n} $ for the parameter \var{n}. In that case
the loop is executed at most $ \maxim\{0,\sym{n}\} $ times (according
to $ \beta((c,d)) $), but the path following the \texttt{if} branch
can be executed at most $\minim\{\maxim\{0,\sym{n}\},3\} $ times. So
the asymptotic complexity for the program is $ \mathcal{O}(\sym{n}) $,
but $ \mathcal{O}(1) $ for the \texttt{if} branch inside the loop.

%\emph{Analysis of Bubble Sort}

In the previous example, we have seen how our algorithm works with
different branches and more bounds for some edges. Another feature
distinguishing our approach from the others is the handling of nested
loops. For the BubbleSort of Figure~\ref{fig:bubble}, our algorithm
computes the bound $\maxim\{0,\sym{n}-\sym{i}-1\}$ for all edges of
the inner loop in the context of the flowgraph induced by the outer
loop. After summarization of the outer loop, the bound changes to
$\maxim\{0,\sym{n}-\kappa_1-1\}$, where $\kappa_1$ is the path counter
of the only backbone of the outer loop. This leads to the
\texttt{else} branch at line~\ref{alg:processLoop:part2:complexCase}
of \texttt{processLoop} and the overall bound
$\sum_{K = 0}^{\maxim(-1,\sym{n}-2)} \maxim(0,\sym{n}-K-1)$. If
$\sym{n}>0$, it can be further simplified to
$\sum_{K = 0}^{\sym{n}-2} \sym{n}-K-1=\frac{\sym{n}\cdot(\sym{n}-1)}{2}$.

\begin{figure}[t] %Bubble Sort
\begin{center}
\begin{tabular}{l}
\texttt{void bubble\_sort(int n, int* A) \{}\\
\texttt{~~for(int i = 0; i < n - 1; i++)}\\
\texttt{~~~~for(int j = 0; j < n - i - 1; j++)}\\
\texttt{~~~~~~if(A[j+1] < A[j]) \{}\\
\texttt{~~~~~~~~int tmp = A[j + 1];}\\
\texttt{~~~~~~~~A[j + 1] = A[j];}\\
\texttt{~~~~~~~~A[j] = tmp;}\\
%\texttt{~~~~~~~~\}\}\}\}}\\
\texttt{~~~~~~\}}\\
\texttt{\}}\\
\end{tabular}
\end{center}
\caption{C code of BubbleSort.}
\label{fig:bubble}
\end{figure}

%}}}
%{{{ Experimental Evaluation

\section{Experimental Evaluation}\label{sec:eval}

We implemented our algorithm in an experimental program analysis tool called
\Looperman. It is built on top of the symbolic execution package
\textsc{Bugst}~\cite{BugstURL} and it intensively uses the SMT solver
\textsc{Z3}~\cite{Z3URL}.

We compare \Looperman with four state-of-the-art loop bound analysis
tools: \Loopus~\cite{Loopus2014}, \KoAT~\cite{KoAT2014},
\PUBS~\cite{PUBS2008}, and \Rank~\cite{Rank2010}.  These freely
available tools were mutually compared before and results from their
elaborate evaluation are also freely
available~\cite{KoAT_EvalData,Loopus_EvalData}. We run \Looperman on
the same benchmarks originally collected from literature.  More
precisely, we use the benchmarks
% \footnote{\url{http://forsyte.at/static/people/sinn/loopus/CAV14}}
translated to C programs by the authors of \Loopus. We ignore programs
with recursive function calls as our tool does not support recursion.
In order to make manual inspection of benchmarks and results from
their analysis manageable, we also ignore all benchmarks associated
with the termination proving tool \textsc{T2}~\cite{T2}. At the end,
we used 199 benchmarks. They are small C programs (ranging from 7 to
451 lines of code, 26 lines in average) containing various kinds of
loops.

% \todo{ Of course, there are more loop bound analysis tools in the
% state-of-the-art \Looperman could be compared with (see section~\ref{sec:rw}).
% Four tools we chose are well known in the area and we believe the comparison 
% with them provides a sufficient insight into performance of the
% presented algorithm. }
% \todo{ We did not compare our tool with other loop bound analysers (see
% section~\ref{sec:rw}) as their input or analysis goals differ significantly from
% our method and the comparison would thus require a lot of manual work. }

We focus on two kinds of bounds: asymptotic complexity bounds for
whole programs and \emph{exact} (meaning non-asymptotic) upper bounds
for individual program loops. The comparison of exact bounds is
restricted to \Looperman and \Loopus as the other tools use input in
different format and (as far as we know) they do not provide mapping
of their bounds to the original C code.

We took the results of \KoAT, \PUBS, and \Rank from the mentioned
experimental evaluation~\cite{KoAT_EvalData}.  In order to obtain the
exact bounds for individual loops (which are not publicly available),
we run \Loopus using a VirtualBox image available on the tool
homepage~\cite{LoopusWeb}. All experiments use 60 seconds
timeout. While \KoAT, \PUBS, and \Rank were executed on a computer
with 6GB of RAM and Intel i7 CPU clocked at 3.07 GHz, \Loopus and
\Looperman run on a machine with 8GB of RAM and Intel i5 CPU clocked
at 2.5GHz. We believe that this difference is not substantial as all
the tools usually either answer very quickly or they fail to determine
any bound. The \Looperman tool (both sources and Windows binaries),
the 199 benchmarks, and all measured data are available at \cite{EVALDATA}.

\subsection{Comparison of asymptotic bounds}\label{sec:evalAsympt}

First we focus on asymptotic complexity bounds for whole programs,
which can be produced by all considered tools. \Looperman derives them
from the upper bounds on edges: each upper bound is transformed into
an asymptotic one (e.g.~$\maxim(0,\sym{a}^2+\sym{b}-1)$ is transformed
into $\mathcal{O}(n^2)$) and maximal asymptotic bound over edges is
then taken.

The results are represented in Table~\ref{tab:benchAsymp}(a). For each tool,
the count of all 199 benchmarks is split into the first three columns. The
columns `correct' and `incorrect' represent counts of inferred
correct and incorrect asymptotic bounds, respectively. The column
`fail' shows counts of benchmarks for which the computation of an
asymptotic bound failed. The column `TO' says how often the fail is
due to timeout. 

We see that \Looperman placed the third after \Loopus and \KoAT in the count
of successfully computed correct asymptotic bounds.  If we consider
incorrect bounds, \Looperman placed the first together with \Loopus. Note
that the incorrect bounds are due to errors in implementation, the
algorithms followed by the tools are in principle sound.
%Nevertheless, in contrast to \KoAT,
%\PUBS, and \Rank, it does not return incorrect bounds.

\begin{table}[t]
\begin{center}
\begin{tabular}{c}
\begin{tabular}{l|cccc}
& correct & incorrect & fail & TO \\
\hline
\Looperman & 129 & 0 & 70 & 5  \\
\Loopus & 162 & 0 & 37 & 0  \\
\KoAT & 140 & 2 & 57&22 \\
\PUBS & 85 & 29 & 85&1 \\
\Rank & 26 & 2 & 171&0 \\[1mm]
\end{tabular}
\\
(a)
\\\\
\setlength{\tabcolsep}{4.5pt}
\begin{tabular}{l|cccc}
& \Loopus & \KoAT & \PUBS & \Rank\\
\hline
succeeds, \textsc{L.} fails & 38 & 36 & 13 & 7 \\
fails, \textsc{L.} succeeds & 5 & 25 & 57 & 110 \\
tighter than \textsc{L.} & 10 & 4 & 2 & 0 \\
less tight than \textsc{L.} & 8 & 9 & 9 & 1 \\
same bounds & 107 & 91 & 61 & 18
\end{tabular}
% \begin{tabular}{r|ccccc}
% & \textbf{Ok} & \textbf{Fail} & \textbf{Better} & \textbf{Worse}  & \textbf{Same} \\
% \hline
% \Loopus & 61 & 2 & 8 & 0 & 94  \\
% \KoAT & 54 & 18 & 3 & 3 & 80 \\
% \PUBS & 28 & 47 & 2 & 3 & 52 \\
% \Rank & 10 & 88 & 0 & 0 & 16
% \end{tabular}
\\
(b)
\end{tabular}
\end{center}
\caption{Numbers of
  correct/incorrect/unknown asymptotic bounds for each tool~(a) and comparison
  of the bounds detected by \Looperman (\textsc{L.}) and by other tools (b).}
\label{tab:benchAsymp}
\end{table}

Table~\ref{tab:benchAsymp}(b) compares results of \Looperman (marked as
\textsc{L.}) relatively to each other tool. We can see the number of cases
where the other tool succeeds to infer a correct bound while \Looperman
fails, or vice versa.  Further, we can see the number of cases where both
tools infer a correct bound and the bound of the other tool is tighter, less
tight, or the same as the \Looperman's bound.

An importation observation from this table is that for each tool there
are benchmarks where the tool fails to compute a correct bound while
\Looperman succeeds. 
%One can conclude that no tool is completely outperformed by \Looperman and also
%\Looperman is not completely outperformed by any other tool (however,
%\Loopus is close).

\subsection{Comparison of exact bounds per loop}\label{sec:evalExact}

Now we compare accuracy of exact upper bounds for individual loops in
benchmarks. Note that \Looperman computes a loop bound as the sum of
bounds on the edges leading from a loop entry node into the loop.

\begin{table}[t]
\begin{center}
\begin{tabular}{c}
\begin{tabular}{l|ccc}
& correct & incorrect & fail \\
\hline
\Looperman & 227 & 0 & 86 \\
\Loopus & 267 & 3 & 43\\
%\multicolumn{4}{c}{\vstrut~}
\end{tabular}
\\
(a)
\\\\
%\textbf{}\setlength{\tabcolsep}{2pt}
\begin{tabular}{l|cc}
& \Looperman & \Loopus \\ \hline
succeeds, the other fails & 11 & 51 \\ 
asymptotically tighter & 16 & 11 \\
tighter (not asymptotically) & 44 & 2
\end{tabular}
% \begin{tabular}{l|ccc}
% & suceeds, the other fails & \textbf{Asym} & \textbf{Exact} \\
% \hline
% \Looperman &  7 & 0 & 33 \\
% \Loopus    & 81 & 7 &  2
% \end{tabular}
\\
(b)
\end{tabular}
\end{center}
\caption{Numbers of correct/incorrect/unknown loop bounds for \Looperman and
  \Loopus (a) and comparison loop bound quality (b).}
\label{tab:benchLoops}
\end{table}

The benchmarks contain 313 loops. Table~\ref{tab:benchLoops}(a) provides for
both tools the numbers of inferred correct and incorrect loop bounds, and the
number of loops for which no bound is inferred.
%
%We see that \Looperman
%produced only correct upper bounds in contrast to \Loopus. Nevertheless,
%\Looperman failed on almost 3 times more loops than \Loopus.
%
Table~\ref{tab:benchLoops}(b) compares quality of the inferred loop
bounds.  It presents the number of loops where one tool produces a
correct loop bounds while the other does not, the number of loops
where one tool provided asymptotically tighter loop bound than the
other, and the number of loops where one tool produces a tighter bound
than the other tool, but the difference is not asymptotic (e.g.~$n$
versus $2n$). To complete the presented data, let us note that both
tools inferred exactly the same bound for 143 loops.

The data in the Table~\ref{tab:benchLoops}(a) show the primary
weakness of \Looperman: It gives up too often.\footnote{\Looperman
  calls the SMT solver Z3 to decide satisfiability of formulae from an
  undecidable logic. However, none of the fails are due to Z3 as it is
  able to solve all queries coming from the benchmarks.} On the other
hand, data in the Table~\ref{tab:benchLoops}(b) show that \Looperman
is able to compute tighter bounds than \Loopus for interesting number
of benchmarks.  
\todo{V tabulkach nesedi cisla: Loopus je lepsi 9x
  asymptoticky na celych programech a jen 7x na cyklech? To je asi
  blbost.}

The results show that \Loopus can infer bounds for slightly more loops
than \Looperman. However, there are also loops bounded by \Looperman
and not by \Loopus. The biggest advantage of \Looperman is definitely
the tightness of its bounds: \Looperman found a tighter bound for
28\% of 216 loops bounded by both tools, while \Loopus found a
tighter bound only for 6\% of these loops.

%\medskip
%All the presented data from our evaluation can be summarised in the following 
%three statements:
%\begin{itemize}
%\item There are loops for which \Looperman computes exact bounds
%while \Loopus fails.
%%
%\item \Looperman gives a higher count of tighter exact upper bounds for loops 
%where both \Looperman and \Loopus succeed.
%%
%\item On the other hand, \Looperman suffers from a lower robustness %scalability
%compared to \Loopus and \KoAT.
%\end{itemize}

% We further note that the reason for although our algorithm builds formulae from
% undecidable logic there was no SAT failure during the whole evaluation
% (both comparisons together).

%}}}
%{{{ Related Work

\section{Related Work}\label{sec:rw}

There are several popular approaches to the computation of upper bounds on
the number of loop iterations. Notable are especially those based on
construction and solving recurrence equations~\cite{PUBS2008, PUBS2007,
  rTuBound2012, ABC}, loop iteration counters~\cite{ABC,Gulwani2009}, and
ranking
functions~\cite{Rank2010,KoAT2014,Loopus2015,Loopus2014,Loopus2011}. Some
tools utilize more than one analysis approach.

Recurrence equations play an important role in \PUBS~\cite{PUBS2008}. It
uses input generated from Java bytecode and computes upper bounds as a
non-recursive representation of a given set of recurrent equations obtained
from other analysers, like~\cite{PUBS2007}. Another representative work is
\textsc{r-TuBound}~\cite{rTuBound2012}. It rewrites multi-path loops into
single-path ones using SMT reasoning. Obtained loops are then translated
into a set of recurrence equations over program variables. Upper bounds are
solutions of the equations. This is done by a pattern-based recurrence
solving algorithm. The analysis implemented in \textsc{ABC}~\cite{ABC}
combines the use of recurrence equations and iteration counters. Authors
focus on nested loops there. An inner loop is instrumented by an artificial
variable increased by one in each iteration (the counter). Recurrence
equations over this variable and regular variables of the loop are
constructed and solved (to get their non-recurrent versions). Bounds on the
artificial variable are obtained by replacing regular variables by their
greatest values in the computed equations.  An advanced use of counters can
be found in \textsc{SPEED}~\cite{Gulwani2009}.  Counters are artificial
variables instrumented into the analysed program initialised to 0 and
incremented by one on back-edges of program loops. A linear invariant
generation tool is then used to compute bounds on the counter variables at
appropriate locations.

In our approach we also use recurrent equations. They are constructed during
symbolic execution and solved in the procedure \texttt{computeSummary} as
functions of path counters. Nevertheless, our computation is simpler than
in the other approaches. And the purpose of equations and their solutions is
also different. We use solutions in the process of construction of necessary
conditions, whose combinations are latter used for inference of upper bounds.

Similarly to \cite{ABC} and \cite{Gulwani2009}, we also use counters. But in
contrast to \cite{ABC}, we introduce a path counter for each acyclic path in
the loop instead of one counter for the whole loop. Our approach is thus
more related to \cite{Gulwani2009}, where counters can also be associated to
individual paths (namely, to back-edges of those paths). Nevertheless, we do
not perform instrumentation of the counter. Another important difference is
that we do not compute bounds on path counters explicitly, while
\cite{Gulwani2009} computes them using an invariant generation tool.

A ranking function was originally used for the termination
proving~\cite{Terminator2006,T2,ARMC,FGPSF,KITTEL}. The ranking function is
usually defined by assigning an expression over program variables to each
program location such that the expression values are always non-negative and
decreasing between every two subsequent visits of the particular
location. It is apparent that a ranking function can be also used for a
computation of upper bounds on the number of iterations of program
loops. \Rank~\cite{Rank2010} reuses results from the termination analysis of
a given program, namely an inferred (global) ranking function, to get an
asymptotic upper bound on length of all possible executions of that
program. \KoAT~\cite{KoAT2014} employs a similar approach, but it uses ranking
functions of already processed loops to compute bounds on values of program
variables. Absolute values of these bounds may in turn improve ranking
functions of subsequent loops. \Loopus~\cite{Loopus2011} transforms an
analysed program in particular locations such that program variables
appearing there represent ranking functions. There are several heuristics
used for this purpose.  Program loops are then summarised which is done per
individual loop paths.  Moreover, a contextualization technique is used to
infer which loop paths can be executed in a sequence.  The approach was
further improved in several directions. Method in~\cite{Loopus2014} merges
nested loops in order to compute bounds for programs, where an inner loop
affects termination of the outer loop. A technique for computing sizes of
variables after loops is introduced in~\cite{Loopus2015}.

Our algorithm does not use ranking functions. However, the passing of
information from a preceding to a subsequent loop we see in~\cite{KoAT2014}
or~\cite{Loopus2015} happens also in our approach. While~\cite{KoAT2014}
and~\cite{Loopus2015} implement this via ranking functions, in our algorithm
it is available due to the use of the symbolic execution. The loop
summarization per individual loop paths presented in~\cite{Loopus2011} is
similar to summarization we perform in the procedure
\texttt{computeSummary}. However, while \cite{Loopus2011} computes summary
as a transitive hull expressed in the domain of a size-change abstraction,
we compute precise symbolic values of variables after the loop. On the other
hand, we do not consider what loop paths can be executed in a sequence. And
finally, in comparison to~\cite{Loopus2014}, we do not merge nested loops.

There are other important techniques computing upper bounds, which
are, however, less related to our work. For instance,
\textsc{SWEET}~\cite{Sweet2006} uses abstract
interpretation~\cite{CousotCousot77} to derive bounds on values of program
variables. The approach further uses a pattern matching for easy and fast
processing of loops with a recognized structure. In \cite{SPEED2010} authors
propose an analysis which is able to compute an upper bound on the number of
visits of a given program location. The program is transformed such that its
execution starts at the location and correctly terminates once the execution
reaches the location again. Loops in the transformed program are summarised
using an abstract interpretation based iterative algorithm into disjunctive
invariants. Upper bounds are computed from the invariants using a
non-iterative proof-rules based technique.
%
% The following reference was added as a reaction to review A from PLDI2016

Another interesting approach for computation of worst-case complexity,
called \texttt{WISE}~\cite{WISE09}, attempts to compute an input for
which the execution of a given program is the longest. From symbolic
executions of all paths up to a given maximal input size there is
inferred a recipe (branch policy generators) which then restricts
symbolic execution of all remaining program paths (for unrestricted
input size) to the longest ones. The worst-case execution input is
then generated from the longest executed path (from its path condition
using an SMT solver).  

There are further approaches
focusing on declarative languages. For example, resource aware
ML~\cite{Hoffmann2012} computes an amortized complexity for recursive
functional programs with inductive data types without a consideration of
upper bounds dependent on integer values. We can further find numerous
techniques for complexity analysis of term rewriting and logic
programming~\cite{AM13,Giesl13,Debray1993,Navas07}.

\section{Future Work}\label{sec:future}

According to results from the evaluation we will primarily focus on
improving scalability of the proposed approach. We have already
analysed problematic benchmarks and we have uncovered a couple of
promising directions for improvements. Here we present three of them.
% which apply to many problematic benchmarks.
%
%  (the complete list of situations in which \Looperman
% failed together with their possible solutions is presented in~\cite{Cad15})
% \todo{ODKAZ SMAZAT}

\begin{enumerate}
\item The loop condition directly implies only inequalities which are non-linear
with respect to the path counters. For example, the analysis of the
  loop % in Figure~\ref{fig:problematic_loops}~(left)
%\begin{center}
 \texttt{while(x*x<n)\,\{\,x=x+1;\,\}}
%\end{center}
%\begin{center}
%\begin{tabular}{l}
%  \texttt{while(x>0)}\\
%  \texttt{~~x=x-t;}
%\end{tabular}
%\end{center}
 leads to the inequality $(\sym{x}^2+\kappa)^2<\sym{n}$. In order to
 compute a bound from this constraint we need a more powerful inequation
 solver.

\item We can add more rules to Algorithm~\ref{alg:compSum} in order to 
  cover more cases of variable value's progression. E.g. in the following loop, the value of
  $\var{x}$ would be $\sym{x} + \sum_{i=0}^{\kappa-1} (\sym{y}+1+i)$.
  % Nevertheless, some examined benchmarks contain loops where
  % control variables follow progressions similar to those presented in
  % the following two examples:
% in Figure~\ref{fig:problematic_loops}~(middle).

  \begin{center}
  \begin{tabular}{l}
    \texttt{while(x<n)~\{}\\
    \texttt{~~y=y+1;}\\
    \texttt{~~x=x+y;}\\
    \texttt{\}}\\
  \end{tabular} 
  \end{center} 
  We plan to extend Algorithm~\ref{alg:compSum} to handle a variety of
  non-linear progressions.

\item There are further loops where arithmetic progressions of control
  variables in outer loops are altered in nested loops as illustrated
  by the following example:
  \begin{center}
  \begin{tabular}{l}
    \texttt{while(x>0)~\{}\\
    \texttt{~~x=x-1;}\\
    \texttt{~~while(y>0)~\{}\\
    \texttt{~~~~y=y-1;~x=x+1;}\\
    \texttt{~~\}}\\
    \texttt{\}}
  \end{tabular}
  \end{center}
  % The inner loop causes a special form of non-linearity of the
  % variable \texttt{x} in the outer loop.
  We plan to deal with such situations according to the solution
  suggested in~\cite{Loopus2014}.
\end{enumerate}
We see no principal issues in the integration of the described
% and other planned
improvements into the algorithm presented here. Nevertheless, some changes in
the control-flow of the Algorithm~\ref{alg:processLoop} and extensions of
Algorithms~\ref{alg:computeBounds} and \ref{alg:compSum} will be necessary.

% \begin{center}
% \begin{figure}[t]
% \begin{tabular}{l}
%   \texttt{while(x>0)~~~while(x<n)~~~while(x>0)}\\
%   \texttt{\{~x=x-t;~\}~~~\{~y=y+1;~~~~~\{~x=x-1;}\\
%   \texttt{~~~~~~~~~~~~~~~x=x+y;~\}~~~~~while(y>0)}\\
%   \texttt{~~~~~~~~~~~~~~~~~~~~~~~~~~~~\{~y=y-1;}\\
%   \texttt{~~~~~~~~~~~~~while(x<n)~~~~~~~x=x+1;~\}\}}\\
%   \texttt{~~~~~~~~~~~~~\{~x=2*x;~\}}\\
% \end{tabular}
% \caption{Schemas of loops in problematic benchmarks.}
% \label{fig:problematic_loops}
% \end{figure}
% \end{center}

%Another direction of our future research is an extension of the
%algorithm to support more features of programming languages like
%writeable arrays, other data types, pointers, procedure calls etc.

% There were already published papers on handling pointers in symbolic
% memory (TODO citace). We can reuse that work and extend the
% Algorithm~\ref{alg:compSum} to include information about pointers in
% our summaries. We further plan to adopt our loop summarisation
% approach for describing effects of recursive calls. }

Another direction of our future research is an extension to compute 
also lower bounds. After that, we plan tu use the method for assertion proving
and invariant generation.

%}}}
%{{{ Conclusion

\section{Conclusion}\label{sec:concl}

We presented an algorithm computing upper bounds for execution counts of
individual instructions of an analysed program during any program run.  The
algorithm is based on symbolic execution and the concept of path
counters. The upper bounds are parametrized by input values of the analysed
program.
% and their computation is based on elimination of
% path counters from expressions resulting from summarised effects of program
% loops.
Evaluation of our experimental tool
\Looperman %implementing our algorithm
shows that our approach is slightly less robust than the leading loop
bound analysis tools \Loopus and \KoAT (i.e.~it infers a bound in less
cases). On the positive side, the loop bounds detected by \Looperman
are often tighter than these found by other tools, which may be a
crucial advantage in some applications including the worst case
execution time (WCET) analysis.

%}}}

\bibliographystyle{abbrv}
\bibliography{loop_bounds_ARXIV}

\end{document}